# A SEMANTICALLY ENRICHED WEB USAGE BASED RECOMMENDATION MODEL

C.Ramesh[1], Dr. K. V. Chalapati Rao[1], Dr. A.Goverdhan[2]

[1]Department of Computer Science and Engineering, CVR College of Engineering,
Ibrahimpatnam, R.R.District, Andhra Pradesh, India.
hmcr.ramesh@gmail.com, chalapatiraokv@gmail.com

[2] JNTUH College of Engineering, Jagityala, Karimnagar District,
Andhra Pradesh, India.
govardhan_cse@yahoo.co.in

## ABSTRACT

*With the rapid growth of internet technologies, Web has become a huge repository of information and keeps growing exponentially under no editorial control. However the human capability to read, access and understand Web content remains constant. This motivated researchers to provide Web personalized online services such as Web recommendations to alleviate the information overload problem and provide tailored Web experiences to the Web users. Recent studies show that Web usage mining has emerged as a popular approach in providing Web personalization. However conventional Web usage based recommender systems are limited in their ability to use the domain knowledge of the Web application. The focus is only on Web usage data. As a consequence the quality of the discovered patterns is low. In this paper, we propose a novel framework integrating semantic information in the Web usage mining process. Sequential Pattern Mining technique is applied over the semantic space to discover the frequent sequential patterns. The frequent navigational patterns are extracted in the form of Ontology instances instead of Web page views and the resultant semantic patterns are used for generating Web page recommendations to the user. Experimental results shown are promising and proved that incorporating semantic information into Web usage mining process can provide us with more interesting patterns which consequently make the recommendation system more functional, smarter and comprehensive.*

## 1. INTRODUCTION AND MOTIVATION

With the rapid growth of internet technologies, Web has become a huge repository of information and keeps growing exponentially under no editorial control. However the human capability to read, access and understand content remains constant. Hence it became more challenging to the Website owners to selectively provide relevant information to the people with diverse needs. Modelling and analyzing Web navigational behavior is helpful in understanding the type of information online user's demand. This motivated researchers to provide Web personalized online services such as Web recommendations to alleviate the information overload problem and provide tailored Web experience to the Web users.

In recent times, Web Usage Mining has emerged as a popular approach in providing Web personalization [1]. However conventional Web usage based recommender systems are limited in their ability to use the domain knowledge of the Web application and their focus is only on Web usage data. As a consequence the quality of the discovered patterns is low. These patterns do not provide explicit insight into the user's underlying interests and preferences, thus limiting the effectiveness of recommendations as well as the ability of the system to interpret and explain the recommendations. Recent studies have suggested that domain knowledge of the Web application in the form of Ontology can play an important role in providing smarter and more comprehensive recommender systems [2]. Hence an increasing effort is needed in defining the Web pages and objects in terms of semantic information in the form of ontologies. The combination of Web Usage Mining and Semantic Web has created a new and fast emerging research area - Semantic Web Mining [3].

Web Usage Mining is defined as an application of data mining techniques on the navigational traces of the users to extract knowledge about their preferences and behavior. The knowledge discovered from Web Usage Mining can be useful in many Web applications such as Web caching, Web prefetching, intelligent online advertisements, in addition to Web personalization [4]. Most of the research efforts in Web personalization correspond to the evolution of extensive research in Web Usage Mining. The main techniques used for modeling personalization systems are clustering pages or user session, association rule generation, sequential pattern generation and Markov models.

In recent years there has been overwhelming attention in employing sequential pattern mining techniques to construct personalization systems. An extensive literature on sequential pattern mining algorithms is available [5] .The aim of Web Usage Mining is to better understand the user's behavior. However, usage behavior is not fully captured in the Web log data. For mining Web usage intelligently, the usage information should be mapped to the semantic space. Web Usage Mining can be enhanced by incorporating semantics into various phases of the Web Usage Mining process.

The Semantic Web is based on a vision of Tim Berners-Lee, the inventor of the WWW. The Semantic Web enriches the WWW by machine – processable information which supports the user in accomplishing his tasks more easily [6]. The vision of a Semantic Web has recently drawn considerable attention both from academic and industrial circles. The idea behind using the semantic Web for generating personalized Web experience is to improve the results of Web mining by exploiting the new semantic structures [2]. As a consequence of the above considerations there is an increasing effort in defining Web pages and objects in terms of semantic information by using ontology.

The aim of this work is to generate frequent sequential patterns enriched with semantic information and the quality of the generated recommendations is evaluated. By using Sequential Pattern Mining algorithm enhanced with semantic information, frequent sequential patterns in terms of ontological instances are found. These results are then used for recommending subsequent pages to the user. To date, the amount of research studies on Web Usage Mining and Semantic Web is very limited, since researchers emphasize primarily on concept hierarchies such as taxonomies. But the expressive power of this form of knowledge representation is limited to is-a relationship [6]. There are studies that aim to generate patterns in terms of semantic information as discussed in [7, 8].

Our recommendation approach uses Web usage analysis powered by an explicit representation of domain knowledge. We hypothesize that such domain knowledge should help increase both the accuracy and interpretability of the discovered patterns. Unlike in previous approaches, in this present work, sequential association rule mining is used to discover frequent sequences.

The rest of the paper is organized as follows: section 2 gives literature review on related work. In section 3, Domain knowledge and Semantic information is presented. In section 4 proposed Architecture is presented. Section 5 gives the experimental results. Finally section 6 includes the conclusions and future scope for enhancements.

## 2. RELATED WORK

The amount of research studies that combines Web Usage Mining and Semantic Web is quite limited. Most of the work shows that integrating the semantics involved in structural links with the Web Usage Mining process can improve the discovered patterns. On the other hand much work was also done in the integration of the content features of site with Web usage mining process [9]. Berendt et al [3], was the first to explore Semantic Web Usage Mining. The authors have elaborated different ways of how the fields of Semantic Web and Web Mining can co-operate. The first part of the work is on extracting semantics from Web page. The second part is on the improvement of Web Usage Mining by using semantics structures in the form of ontology. Subsequently the authors sketched out the benefits of combining Semantic Web and Web Mining. They further elaborated a process for learning ontologies by mining the Web. They emphasized that by constructing a pattern space over ontology, navigation primitives and Web Mining methods one can find patterns which follows a semantic approach.

The studies presented in [10], extracted domain level objects from user sessions and created a user profile for each user by aggregating these objects according to their weights and a merge function. It is assumed that there already exists a domain level ontology for the Website and merge functions have already been defined on every attribute of objects. Another related work given in [6], investigated the notion of semantics and ontologies in detail and discussed the possible ways in which the Semantic Web can improve the results of Web Usage Mining by exploiting the Semantic structures such as ontologies and how Web Usage Mining can help to build up the Semantic Web. The paper discusses how Web Usage Mining can be utilized to learn ontological structures and classify the instances. The studies presented in [11, 12] integrated Content Mining into Web Usage Mining, for effective personalization.

In a recent work [13], Nasraoui et al, proposed a Web Usage Mining Framework for mining evolving User Profiles of dynamic Web sites by exploiting the external ontology, used for mapping and relating dynamic Web pages. Eirinaki et al [14], presented a system, SEWeP which integrates the Web usage logs with the semantics of Web site's content to improve the personalization. The innovative feature of the architecture was C-logs, an extended form of Web usage log which encapsulates the site semantics. But the framework was limited only to concept hierarchy.

Amit Bose et al, [15] proposed a framework for personalization combining usage information and domain knowledge based on ideas from bioinformatics and information retrieval. Unlike our model, these works do not integrate the domain knowledge of the Web application in all phases of Web usage Mining.

In the present paper OntoSPM algorithm is described which extends the algorithm discussed in the literature [5], by the use of richer ontology, to prune the candidate frequent sequences and thereby reduce the number of database scans.

## 3. DOMAIN KNOWLEDGE AND SEMANTIC INFORMATION

The classical workflow of the Web usage mining process is extended with new steps, by integrating domain knowledge in the form of ontology. In this work, it is assumed that domain knowledge is available in the form of domain ontology provided by the ontology engineer during the design of the Web site.

An ontology is an explicit specification of a conceptualization i.e., a specification of an abstract, simplified view of a domain. Formally and in accordance with [16], a core ontology is represented as a 5-tuple $O := \{C, R, H^c, rel, A\}$ where $C$ is a set of concepts which represent the entities in the ontology domain; $R$ is a set of relations, defined among the concepts; $H^c$ is a taxonomy or concept hierarchy which defines the is-a relation among concepts; *rel* element corresponds to a function, *rel*: $R \rightarrow C \; x \; C$ that specifies the relations on *R*. *A* is a set of axioms usually expressed in a logical language. For example the set C contains Product, Purchase, Supplier, and Warehouse as some of the concepts of the domain ontology considered in our Model.

## 4. PROPOSED ARCHITECTURE

We have distinguished two stages in the whole process – i) offline tasks that includes data preprocessing and cleaning followed by Sequential Association rule mining, ii) online tasks that concern the generation of recommendations as shown in the fig.1, which outlines the proposed architecture.

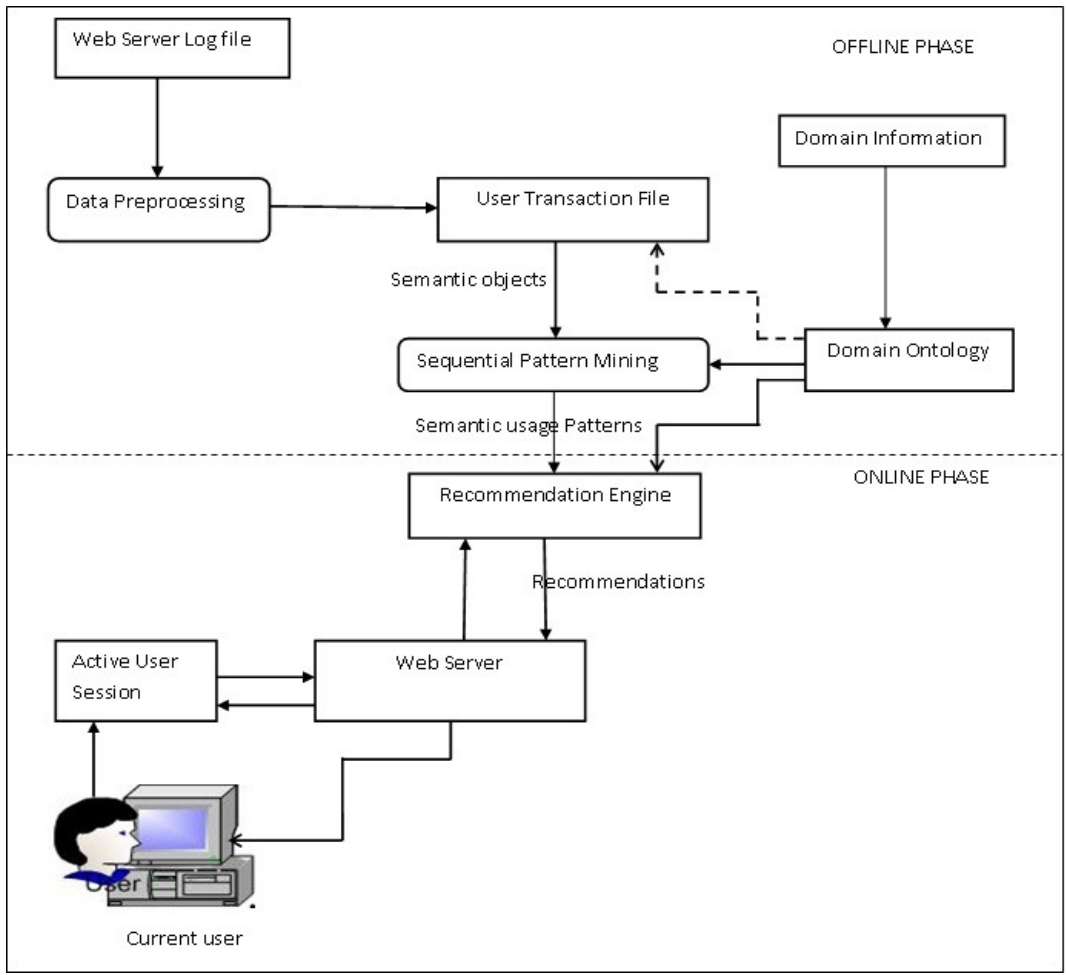

Figure.1 Proposed Recommendation Model incorporating Semantic information in Web Usage Mining process

## 4.1. OFFLINE PHASE

### 4.1.1. Data Preprocessing and Cleaning

The preprocessing phase is the first component in the architecture. Web server log file, which is the main source of input, generally contains noisy and irrelevant data. Preprocessing phase consists of data cleaning, user's identification and session identification tasks. During preprocessing Web server log files are pruned to remove irrelevant requests such as non responded requests and requests made by software agents such as Web crawlers and search engines.

Objects representing products in our e-commerce Web site (http://www.orderzone.com) and which are dealt with in the mining process are considered as instances of concepts (also called classes). The domain ontology was developed using a standard framework, Protégé and OWL[17] was used as ontology representation language. Each Web page is annotated with semantic information during the development of the Website thus showing which ontology class it is an instance of. The cleaned and filtered Web log file is passed to ontology based Web log parser and all the ontology instances represented by the Web pages are extracted converting the Web log to a sequence of semantic objects. The preprocessing tasks results in aggregate structures such as user transaction file, containing semantic objects where each object is represented as tuple $<pg, inst_i>$, where $pg$ represents the Web page which contains the object/product, usually an URL address of the page, and $inst_i$ is an instance of a class c $\epsilon$ $C$, from provided ontology $O$, where $i$ is an index for an enumeration of the objects in the sequence, from the Web sequence database being mined.

### 4.1.2. Sequential Pattern Mining

In the present work an algorithm OntoSPM , a variation of Apriori algorithm, enhanced by semantic information for generating frequent sequences [5]  and Apriori_Join() for candidate generation is described. OntoSPM generates semantically rich frequent sequences. Apriori_Join procedure uses Semantic distance to prune the candidate sequences, such that if the semantic distance between two (k-1) sequences is more than an allowed maximum semantic distance $\delta$ , then the candidate k-sequence is pruned from the search space without the need for support counting. OntoSPM algorithm and Apriori_Join procedure is given below. The definitions and terms used in this algorithm are from [18].

Algorithm1 Ontology Based Sequential Pattern Mining
*OntoSPM ($S, M, \delta, min\_sup$)*
**Input:** Sequence database *S, /\* sequence of semantic objects \*/*
        Semantic distance Matrix *M,*
        Maximum semantic distance *$\delta$, /\* $\delta$: Maximum Semantic distance to prune the search*
                                             *Space \*/*
        Minimum support *min_sup*
**Output:** Semantic-rich Frequent Sequences
**Algorithm:**
1: Scan database S to find the set of frequent 1-sequence $L_1$ = {$s_1$, $s_2$, …$s_n$}
2: k=1,
3: $C_1$=$L_1$
4: for (k=2; $L_{k-1} \neq \emptyset$; k++)
5:     for $L_{k-1} \bowtie L_{k-1}$ do       /\* $L_{k-1}$ join $L_{k-1}$ \*/
6:           There exist $s_i$,$s_j$ such that $s_i$, $s_j \epsilon L_{k-1}$
7:           $C_{k\leftarrow} C_k \bigcup$ Apriori_Join($L_{k-1}$,$M$, $\delta$)
8:           end for
9:       $L_{k=}$ {c $\epsilon$ $C_k$ | c.support $\geq$ min_sup}
10:  end for
11:  return L=$\bigcup_k L_k$
    end

*Function Apriori_join () implementation is a variation of the join procedure of the sequential pattern mining algorithm (GSP).*

Function *Apriori_join* ($L_{k-1}$, *M, δ*)

$C_1$=Ø

for all P,Q ϵ $L_{k-1}$

        with P = { $i_1, i_2, \ldots i_{k-2}, i_{k-1}$}

          and Q = {$i_1, i_2, \ldots i_{k-2}, i'_{k-1}$}

           and $D$ ($i_{k-1}, i'_{k-1}$) ≤ *δ*   /* $D$($i_{k-1}, i'_{k-1})$: defines the semantic distance between $i_{k-1}, i'_{k-1}$ */

                                            /* *δ*: Maximum Semantic Distance -

                                                  Maximum allowed semantic distance between any two semantic objects. *δ* is a user defined value and can be determined as specified in [18]. $D$($i_{k-1}, i'_{k-1}$) is derived from *M* */

                        c = {$i_1, i_2, \ldots i_{k-1}, i'_{k-1}$}

$C_k \leftarrow C_k \cup$ { c }

return $C_k$.

## 4.2. ONLINE RECOMMENDATION PHASE

The aim of a recommender system is to determine which Web pages are more likely to be accessed by the user in the future. In this phase active user's navigation history is compared with the discovered Sequential Association rules in order to recommend a new page or pages to the user in real time. Generally not all the items in the active session path are taken into account while making a recommendation. A very earlier page that the user visited is less likely to affect the next page since users generally make the decision about what to click by the most recent pages. Therefore the concept of window count is introduced. Window count parameter '*n*' defines the maximum number of previous page visits to be used while recommending a new page. Since the association rules are in the form of ontology individuals, the user's navigational history is converted into the sequence of ontology instances. Then the semantic rich association rules and user navigation history are joined in order to produce recommendations.

In the recommendation phase, in the first instance, the most recently navigated item is taken as the search pattern. All the semantic–rich association rules are scanned and the association rules whose antecedent part is equal to the search pattern are added to the recommendation set. This step iterates window count times and at each iteration, the search pattern is extended by one item. The recommendation set constitutes the set of semantic rich association rules sorted in the decreasing order of their confidence. After constructing the recommendation set, the page recommendation commences. Semantic distance between objects [18] is taken into consideration to solve the ambiguity problem. For instance consider the following two semantic rich association rules and

AB→ C

AB→D

Where A, B, C, D are semantic objects. If semantic distance (B,D) < semantic distance(B,C) meaning that D is semantically closer to B than C is, then recommendation engine will prefer D over C and the page(s) representing product D will be recommended. Such capability is not provided by regular association rules. The consequent part of the rule contains ontology individuals; therefore the instances should be converted to the real Web objects. The Web pages for the Web objects present in the recommendation set are recommended.

## 5. EXPERIMENTAL RESULTS AND EVALUATION METRICS

For the purpose of this experiment we have utilized the Web server's log file of an e-commerce Web site "OrderZone "(http://www.orderzone.com). This Web site contains 300 products in 60 categories. We considered a log file corresponding to a two weeks period. This has approximately size of 230 MB. A moderate amount of traffic of an average 2500 visits and 6000 page-views per day is measured on the site. After preprocessing the log file, 1600 unique sessions with an average of 4 page views per session were obtained.  The ontology model of the domain includes the concepts Product, Supplier, ProductCategory, Order, Purchase, warehouse. The Algorithm was run with MinimumSupport = 0.01 and Maximum Semantic Distance $\delta$ =10.

The Evaluation method of this work is based on the metrics introduced in [19]. The effectiveness of the recommendation is measured in terms of precision and coverage. 10-fold cross-validation is performed on the data set. Each transaction $t$ in the test set is divided into two parts. The first part is the first $n$ items in $t$ for recommendation generation; $n$ is called the *window count*. The other part, which is denoted as *Eval*, is the remaining portion of $t$ to evaluate the recommendation. Once the recommendation phase produces a set of page views, which is denoted as *Rec,* the set is compared with *Eval* page views.

Precision is defined as the proportion of the number of relevant recommendations to the number of all recommendations.

$$precision = \frac{|\ Rec \cap Eval\ |}{|Rec\ |}$$

Coverage measures the ability of the recommendation system to produce all the page views that are likely to be visited by the user.

$$coverage = \frac{|Rec \bigcap Eval|}{|Eval|}$$

We performed experiments with recommendation thresholds ranging from 01.to 1.0 .The results of these experiments are presented below.

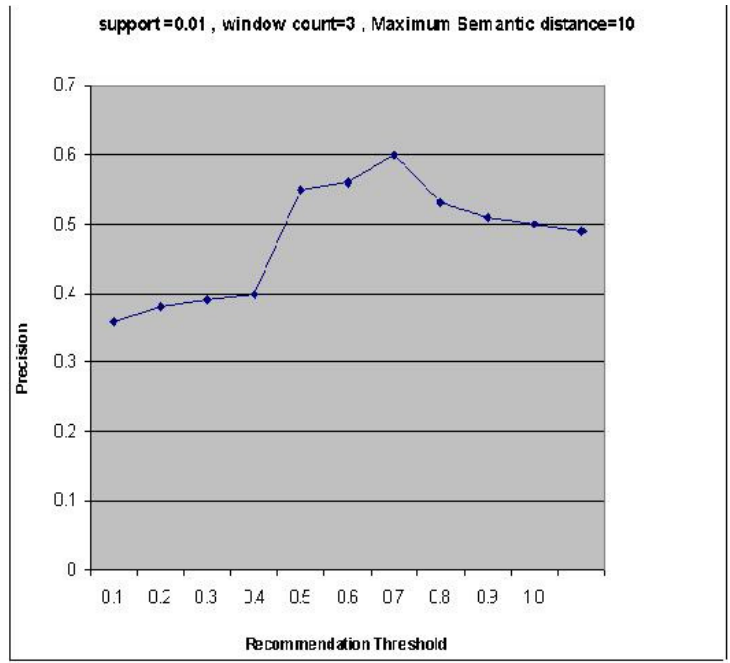

Figure 2. Precision of the Recommendation Model

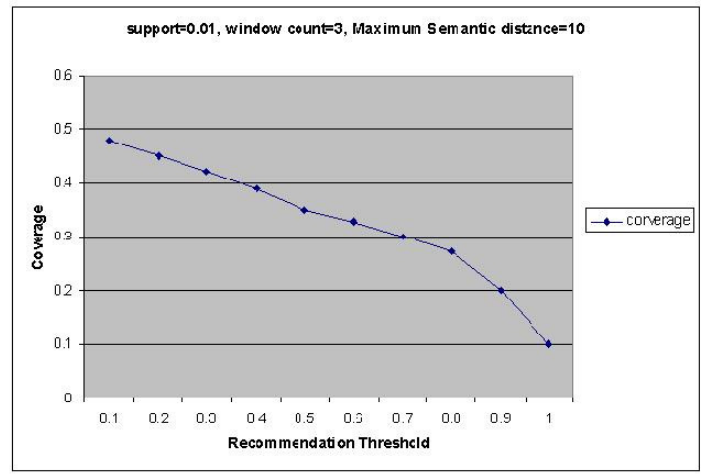

Figure 3. Coverage of the Recommendation Model

## 6. CONCLUSIONS AND FUTURE WORK

Patterns generated by conventional Web Usage Mining methods do not provide explicit insight into the user's underlying interest and preferences. Hence there is a need to incorporate semantic information in web usage model to understand web user's navigational behavior at conceptual level. This motivated us to propose the semantically enriched web usage model. The proposed work integrates domain knowledge in the form of ontology in all the phases of Web Usage Mining process. The generated patterns are in terms of ontology instances instead of Web page addresses. Such patterns can extract the semantic relatedness between the visited Web pages. The discovered Semantic rich sequential association rules form the core knowledge of the recommendation engine of the proposed model. Compared with the conventional Web usage based recommendation system, our proposed model shows promising results.

The present work can be extended in the following directions:
1) Elaborating the use of semantically enhanced patterns for automatic or semi automatic Web site adaptation and link restructuring.
2) Incorporating domain knowledge of the web application in the form of ontology in pattern - growth sequential pattern mining algorithms.

## REFERENCES

[1]     B. Mobasher, Robert Cooley and Jaideep Srivastava, (2000) "Automatic personalization based on Web usage mining", *Communications of the ACM, 43(8)*, pp. 142-151.

[2]     Honghua Dai and Bamshad Mobasher, (2005) "Integrating Semantic Knowledge with Web Usage Mining for Personalization", *Web Mining: Applications and Techniques, Anthony scime (eds.)*, IRM Press, Idea Group Publishing, 2005.

[3]     B.Berendt, A. Hotho and G. Stumme, (2002) "Towards Semantic Web Mining", *Horrocks, I., Hendler, J. (eds.) ISWC 2002, LNCS*, Vol. 2342, pp. 267-278, Springer, Heidelberg (2002).

[4]     J. Srivastava, R. Cooley, M. Deshpande and P. Tan, (2000) "Web usage mining: Discovery and applications of usage patterns from Web data", *SIGKDD Explorations*, Vol. 1, No. 2, pp. 12-23, 2000.

[5]     R. Agrawal and R. Srikant, (1995) "Mining Sequential Patterns", *Proceedings. of the 11th Int'l Conference on Data Engineering (ICDE- 95)*, pp. 3-14, March 1995.

[6]      G.Stumme, B. Berendt and A Hotho, (2004) " Usage Mining for and on the Semantic Web", *Data Mining: Next Generation Challenges and Future Directions*, pp. 461-480, AAAI/MIT Press.

[7]     M.Eirinaki, D. Mavroeidis, George Tsatsaronis and M. Vazirgiannis, (2006) "Introducing Semantics in Web Personalization: The Role of Ontologies, Semantics, Web and Mining*", Springer-Verlag*, Berlin Heidelberg, 2006, pp. 147-162.

[8]     L. Wei and S. Lei, (2009) "Integrated Recommender Systems Based on Ontology and Usage Mining", *Active Media Technologies, 5820, Springer-Verlag*, Berlin Heidelberg, pp. 114-125, 2009.

[9]     M. Eirinaki, C.Lampos, S. Paulakis and M. Vazirgiannis, (2004) "Web Personalization Integrating Content Semantics and Navigational Patterns", *Proceedings of WIDM'04*, USA.

[10]    H. Dai and B. Mobasher, (2002) "Using Ontologies to discover domain- level Web Usage profiles", *Proc. of the 2nd Semantic Web Mining Workshop at ECML/PKDD 2002*, Helsinki, Finland, 2002.

[11]    B. Mobasher, H. Dai, T. Luo, Y. Sun and J. Zhu, (2000) "Integrating Web Usage and Content Mining for More Effective Personalization*", Proc. of the International Conference on E-Commerce and Web Technologies (ECWeb2000)*, pp. 165-176, Greenwich, UK, 2000.

[12]    B. Mobasher, H. Dai, T. Luo, Y. Sun and J. Zhu, (2009) " Integrating Web Content Mining into Web Usage Mining for finding patterns and predicting users' behavior", *International Journal of Information Science and Management,* Vol.7, No.1, January/June 2009.

[13]    O.Nasraoui, Maha Soliman,  Esin Saka,  Antonio Badia and  Richard Germain, (2008) " A Web Usage Mining Framework for mining evolving user profiles in dynamic Web sites", *IEEE Trans. Knowl. Data Eng.* Vol. 20, No. 2, pp. 202-215.

[14]    M. Eirinaki, M. Vazirgiannis and I. Varlamis, (2003) "SEWeP: Using Site Semantics and a Taxonomy to Enhance the Web Personalization Process", *Proc. of the 9th SIGKDD Conf,* 2003.

[15]    Amit Bose, Kalyan Beemanapalli, Jaideep Srivastava and Sigal sahar, (2006) "Incorporating Concept hierarchies into Usage Mining Based Recommendations", *Proceedings of WEBKDD'06*, Pennsylvania.

[16]    G.Stumme , A. Hotho and B. Berendt, (2006) "Semantic Web Mining: State of the art and future directions", *Journal of Web Semantics: Science ,Services and Agents on the World Wide Web*, Vol. 4, No. 2, pp. 124-143.

[17]    http://www.w3.org/TR/owl-features/

[18]    Nizar Mabroukeh and C.I. Ezeife, (2009) "Using domain ontology for Semantic Web usage mining and next page prediction", *Proceedings of the 18th ACM Conference on Information and Knowledge Management (CIKM)*, Hong Kong, November 2-6, 2009,  pp. 1677-1680.

[19]    Miki Nakagawa and Bamshad Mobasher, (2003)"Impact of site characteristics on Recommendation Models Based on Association Rules and Sequential Patterns", *Proceedings of the IJCAI'03 Workshop on Intelligent Techniques for Web Personalization*, Acapulco, Mexico, August 2003.

**Authors**

Mr. C. Ramesh is working as Associate professor in Computer Science and Engineering Department at CVR College of Engineering. He is pursuing his PhD from Jawaharlal Nehru Technological University, Hyderabad. He received his B.E in Computer Science and Engineering from Osmania University and M.Tech from Jawaharlal Nehru Technological University. His areas of interest include Databases, Data Mining, Semantic Web Mining, Web Usage Mining, and Social Networks.

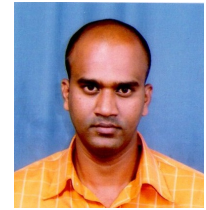

Dr K.V. Chalapati Rao is a Professor of Computer Science & Engg., and Dean, Academics at CVR College of Engineering. Prior to joining the CVR, he served Osmania University as a Professor & Head, Department of CSE and Dean of Engineering. After obtaining his PhD, Dr. Rao joined Electronics Corporation of India Limited and worked in various capacities for 16 years, before joining the Osmania University. He guided number of PhD scholars in areas of Real time systems, Operating Systems, Software Engineering, Distributed Systems, Knowledge and Data Engineering.

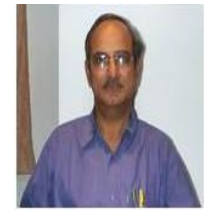

Dr.A.Govardhan: received Ph.D. degree in Computer Science and Engineering from Jawaharlal Nehru Technological University in 2003, M.Tech from Jawaharlal Nehru University in 1994 and B.E from Osmania University in 1992. He is working as a Principal of Jawaharlal Nehru Technological University, Jagitial. He has published around 108 papers in various national and international Journals/conferences. His research of interest includes Databases, Data Warehousing & Mining, Information Retrieval, Computer Networks, Image Processing, Software Engineering, Search Engines and Object Oriented Technologies